\documentclass[%
 reprint,
 amsmath,amssymb,
 aps,
]{revtex4-2}

\usepackage{graphicx,epstopdf}
\usepackage{dcolumn}
\usepackage{bm}
\usepackage{hyperref}

\usepackage{xcolor}
\usepackage{soul}
\usepackage{tikz} 
\usepackage{lipsum}
\usepackage{amsmath}
\usepackage{amsfonts}
\usepackage{verbatim}
\usepackage{float}
\usepackage{caption}
\usepackage[caption=false]{subfig}
\usepackage[utf8]{inputenc}

\hypersetup{
    bookmarks=true,         
    unicode=false,          
    pdftoolbar=true,        
    pdfmenubar=true,        
    pdffitwindow=false,     
    pdftitle={Certificate},    
    pdfauthor={Dr. Harish Kumar},     
    pdfsubject={TEQIP certificates},   
    pdfcreator={Dr. Harish Kumar},   
    pdfproducer={},  
    pdfkeywords={Certificates,} {TEQIP} {Participation}, 
    pdfnewwindow=true,      
    colorlinks=false,       
    linkcolor=red,          
    citecolor=green,        
    filecolor=magenta,      
    urlcolor=cyan           
}

\begin{document}

\title{Time-heterogeneity of the F\"{o}rster Radius from Dipole Orientational Dynamics Impacts Single-Molecule FRET Experiments}

\author{David Frost}
 \affiliation{School of Mathematical and Statistical Sciences, Clemson University.}
 \author{Keisha Cook}%
 \email{keisha@clemson.edu}
 \thanks{Corresponding author}
\affiliation{School of Mathematical and Statistical Sciences, Clemson University.}
\author{Hugo Sanabria}
 \email{hsanabr@clemson.edu}
\affiliation{
 Department of Physics and Astronomy, Clemson University.\\
}

\date{\today}

\begin{abstract}
F\"{o}rster resonance energy transfer (FRET) is a quantum mechanical phenomenon involving the non-radiative transfer of energy between coupled electric dipoles. Due to the strong dependence of FRET on the distance between the dipoles, it is frequently used as a ``molecular ruler" in biology, chemistry, and physics. This is done by placing dipolar molecules called dyes on molecules of interest. In time-resolved confocal single-molecule FRET (smFRET) experiments, the joint distribution of the FRET efficiency and the donor fluorescence lifetime can reveal underlying molecular conformational dynamics via deviation from their theoretical F\"{o}rster relationship. This deviation is referred to as a dynamic shift. Quantifying the dynamic shift caused by the motion of the fluorescent dyes is essential to decoupling the dynamics of the studied molecules and the dyes. We develop novel Langevin models for the dye linker dynamics, including rotational dynamics, based on first principle physics and proper dye linker chemistry to match accessible volumes predicted by molecular dynamics simulations. By simulating the dyes' stochastic translational and rotational dynamics, we show that the observed dynamic shift can largely be attributed to the mutual orientational dynamics of the electric dipole moments associated with the dyes, not their accessible volume. Our models provide the most up-to-date and accurate estimation of FRET.

\end{abstract}


\maketitle

\section{\label{sec:intro}Introduction}

F\"{o}rster resonance energy transfer (FRET)\cite{clegg1996fluorescence, Clegg2009} experiments are commonly used in biochemistry and biophysics to measure distances at the molecular level to resolve conformational states.\cite{stryer, smFRET_1, roy2008practical, Walt2013, hellenkamp2018precision} This is done by exploiting the physical mechanism of FRET wherein energy is transferred non-radiatively between a {\it{donor}} and an {\it{acceptor}} fluorophore due to electric dipole coupling. For simplicity, we refer to these fluorophores as dyes. In FRET, the rate of energy transfer depends on the distance between the dipoles to the sixth power, as well as an orientational factor and several other time-independent factors.\cite{clegg1996fluorescence, roy2008practical, ha2024fluorescence, knox2012forster, smFRETreview} Due to this strong dependence on distance, FRET is often used as a {\it{molecular ruler}} \cite{stryer,schuler2008protein}. However, due to the many distance-related interpretations in FRET and potential pitfalls, assignments of FRET derived distances to specific structural properties is not trivial \cite{Ingargiola2018, peulen2017combining, Saurabh2024, gotz2024reply, ha2024fluorescence}. For example, measuring distances within single molecules (smFRET) is a well-established concept in structural biology.\cite{craggs2012six,lerner2021fret,dimura2020automated, schuler2013single, schuler2008protein, nagy2015complete, torella2011identifying, hoffmann2011quantifying, gopich2007single, chung2011extracting, hofmann2012polymer, doi:10.1021/ar0401451, doi:10.1126/science.1086911, HA200178, doi:10.1073/pnas.96.3.893, lerner2018toward, Multistate_2, agam2023reliability, WeissEtAl2006, Weiss2000, schuler2013single, dimura2020automated} Hence, accurately understanding FRET measurements is crucial for interpreting experimental data and forming concrete hypotheses about the conformational dynamics of the molecule of interest. This understanding is essential for developing accurate biosensors, elucidating signal transduction pathways, studying biomolecular interactions, advancing drug development, and creating fluorescence spectroscopic toolkits. \cite{sanabria2020resolving, ShresthaEtAl2015}

FRET can be measured via photon counting or time-resolved methods.\cite{lakowicz} Previously, we and others demonstrated \cite{Multistate_1,Multistate_2, kalinin2010origin} that the joint distribution between FRET efficiency—the proportion of excited donor molecules transferring energy to acceptor molecules—and fluorescence lifetime—the time a molecule remains in the excited state—provides extensive information on the structural dynamics of the underlying molecule. Thus, we introduced the term "dynamic shift" to account for deviations from the ideal F\"{o}rster relationship\cite{Multistate_1}, with multiple exampled on how to the dynamic shift concept helped understand the structural dynamics of biomolecules.\cite{lerner2018toward, Multistate_2, agam2023reliability, hamilton2022fuzzy, krishnamohan2023coevolution, torella2011identifying} Due to thermal noise, the dye position fluctuates randomly, altering the joint FRET-lifetime distribution.\cite{sindbert2011accurate, kalinin2010origin} Therefore, it is crucial to understand how these stochastic fluctuations influence the resulting FRET measurements.\cite{doi:10.1021/jp075255e, doi:10.1021/jp9634518, ha1996probing, deniz1999single, gopich2012theory} A comprehensive understanding of the dynamic shift induced by dye dynamics is needed to achieve accurate estimates of structural dynamics.\cite{doi:10.1126/science.265.5170.361, graen2014amber}   

Current dye models in the literature lead to significant uncertainties in FRET measurements.\cite{sindbert2011accurate,kalinin2010origin} Moreover, the available models are either simplistic\cite{kinosita1977theory} or require high computational costs.\cite{nagy2015complete, holmstrom2018accurate, graen2014amber, peng2019electrostatic} For example, all-atom molecular dynamics (MD) simulations are often conducted on a sample-by-sample basis to calculate the accessible volume of the dyes, thereby estimating the range of motion of the dye and the labeled molecule while providing a method for uncertainty quantification in the FRET-lifetime distribution.\cite{graen2014amber, yanez2018identifying, agam2023reliability, WANG20142280} However, a key critique of this method is that MD simulations do not last for the entire sampling process of typical smFRET experiments.\cite{graen2014amber} Additionally, MD simulations are designed to approximate the equilibrium distribution of the dye motion, ignoring any time heterogeneity in the energy transfer process itself. Consequently, the full range of dye motion during a sample period is not captured.

Here, we present a semi-analytic model for fluorescent dye motion that addresses the following questions: First, will an isotropic Gaussian process provide a correct model? Second, what is the impact of linker length? Third, what is the role of dipole motion in FRET measurements? We answer these questions in the context of simulated smFRET experiments. Our results challenge the previous assumption that the dynamic shift induced by the dyes is solely due to the accessible volume of the dye (i.e., translational motion).\cite{Multistate_1,Multistate_2} This work demonstrates that the dynamic shift actually depends on the full state dynamics of the dyes, including translational and orientational movements. Therefore, to reduce uncertainty in FRET-derived distances or determine the structural dynamics of biomolecules, we can decouple the individual dye dynamics from the dynamics of the molecule under study by their dynamic shift signature.

\subsection{\label{sec:Exp_FRET} Confocal smFRET}

A time-resolved confocal smFRET experiment uses the FRET phenomena by attaching fluorescent dyes to a molecule of interest by continuously exciting the donor molecule and measuring the resulting emitted light; one may estimate the FRET efficiency and, from that, estimate the distance between the molecules. 

The estimation of FRET efficiency is done in two ways deemed \emph{intensity} based FRET and \emph{lifetime} based FRET.\cite{george, roy2008practical} Both methods may be used in time-resolved confocal smFRET, and indeed, the subject of this paper concerns the relationship between the two. 

The experiment is carried out by diluting the sample of interest so that, on average, less than one molecule of interest lies within the confocal volume of the microscope.\cite{george, roy2008practical} By freely diffusing through the confocal volume, the attached donor dye becomes repeatedly excited, and the resulting fluorescence is measured. The random amount of time spent within the confocal volume is termed a burst time due to the burst of photons seen in the measurement process.  Moreover, by recording the time between the laser pulse and measurement, one obtains the \textit{lifetime} measurement \cite{george}. The lifetime measures how long the molecule remains excited, a quantity that will change in the presence of other molecules. 

A \emph{burst time} is defined as the amount of time the molecule diffuses through the confocal volume. During this time, the inter-photon arrival time decreases sharply from the background level. Through this stochastic time frame, measurements are made and one obtains a sample of lifetimes and photons by which the FRET efficiency and lifetime distribution may be estimated. By collecting multiple burst samples, one obtains the joint FRET-lifetime distribution. During each burst, the donor molecule repeatedly enters an excited state wherein it may fluoresce, transfer energy via the FRET mechanism, or relax due to some other relaxation pathway. We refer to the event of excitation and the following process dictating the observation as an \emph{excitation event}. In simple terms, a burst time will give a data point in the FRET-lifetime distribution and represents \emph{multiple} photons. Whereas an excitation event gives the color and lifetime of \emph{one} photon.

\subsection{\label{sec:fret}FRET Model}

Consider two completely static dyes with normalized dipole moments $\hat{\mu}_A \in S^2$ and $\hat{\mu}_D \in S^2$ for the acceptor and donor, respectively. Further, let the inter-dye displacement vector be $r \in \mathbb{R}^3$. The energy transfer rate is defined in Eq.~(\ref{eq:KET}),
\begin{equation}
    k_{ET}(r) = k_D\bigg(\frac{R_0}{\|r\|}\bigg)^6,
    \label{eq:KET}
\end{equation}
 where $k_D$ is the fluorescence decay rate for the donor dye, and $R_0$ is the distance at which the energy transfer efficiency is 0.5
 \cite{Clegg2009,lerner2021fret, roy2008practical, gopich2012theory}. Note that the energy transfer rate increases steeply as the distance decreases and inversely as the distance increases. However, no matter the distance, no energy transfer can happen if $\hat{\mu}_D, \hat{\mu}_A$, and $\hat{r}$ are mutually orthogonal. \cite{vandermeer2020kappaphobia, van2002kappa} The F\"{o}rster radius can be written as $R_0^6(t) = C\kappa^2(t)$ where $C$ is a constant depending on the environment surrounding the dye. The parameter $\kappa^2(t)$ is the dipole orientational factor \begin{equation}
     \kappa(t) = (\hat{\mu}_D(t) \cdot \hat{\mu}_A(t)) - 3(\hat{r}\cdot \hat{\mu}_D(t))(\hat{r}\cdot \hat{\mu}_A(t))
     \label{eq:kappa}
 \end{equation} with $\hat{r} = \frac{r}{\|r\|}$.\cite{dale1979orientational} . Since the dipole moments are known to reorient on timescales faster than the energy exchange rate\cite{van2002kappa, munoz2009fretting}, kappa square is treated as time-dependent. This is in contrast to previous models wherein the dipole moment is chosen from the equilibrium distribution of the rotational diffusion. \cite{eilert2018complete, gopich2012theory, Jares-Jovin} In this model, the initial distribution of the dipoles is chosen according to the equilibrium distribution, but rotational processes evolve during the energy transfer. This is vital because the FRET efficiency cannot be evaluated in terms of an evaluation of the energy transfer rate at a specific time but rather as dependent on the history of the $\kappa^2$ process using the fact that the transfer times at a time $T >0$ of a non-homogeneous CTMC are exponential with rate $\int_0^T k(s)ds$ \cite{stroock2004introduction, Durrett, Ross, van1992uniformization} therefore the FRET efficiency at time T is given by \begin{equation}
     \mathcal{E}(T) = \frac{\int_0^T k_{ET}(s)ds}{ \int_0^T k_{ET}(s) ds + k_DT}.
     \label{eq:time-inhomogeneousFRET}
 \end{equation} In this way, the FRET efficiency process, $\mathcal{E}(t)$, is non-Markovian. It is important to note that each excitation event's fluorescence process will still be Markovian. As noted in \cite{eilert2018complete}, the inter-arrival time for the photon count process need not be exponentially distributed.\cite{Yakovlev2005InterarrivalTD} Therefore, the photon arrival process cannot be seen as a time-homogeneous Poisson process in contrast to previous common assumptions.\cite{smFRET_1, smFRET_2, smFRET_3, gopich2007single,nir2006shot} Depending on the rate at which the dyes reorient, each vector may be treated as uniformly distributed on the unit sphere or a cone.\cite{eilert2018complete} In this case the average value of $\kappa^2$ is given by $\frac{2}{3}$.\cite{clegg1996fluorescence} This is referred to as the dynamic averaging regime \cite{dale1979orientational,van2002kappa, eilert2018complete}. 

\begin{figure}
    \centering
    \includegraphics[width=.45\textwidth]{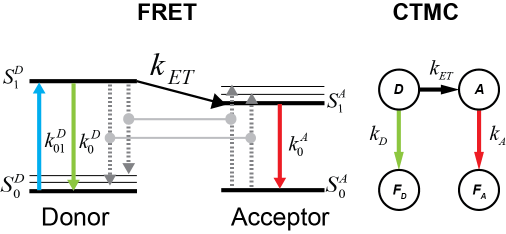}
    \caption{Representation of FRET by CTMC. Comparison of the CTMC states with the Jablonski diagram for the FRET process.}
    \label{fig:Jablonski_CTMC}
\end{figure}

Using the fact that exponential random variables can well model fluorescence times \cite{lakowicz} and accounting for the time dependence of the F\"{o}rster radius on $\kappa^2$ the energy transfer process in FRET is modeled as a time-inhomogeneous continuous-time Markov chain (CTMC)\cite{Ross, Durrett}, illustrated in Figure~\ref{fig:Jablonski_CTMC}, with rate matrix defined in (\ref{ratematrix});
\begin{equation}
    Q(t) = \begin{pmatrix}
    -(k_D + k_{ET}(t)) & k_{ET}(t) & k_D & 0 \\ 
    0 & -k_A & 0 & k_A \\ 0 & 0 & 0 & 0 \\ 0 & 0 & 0 & 0
    \end{pmatrix},
    \label{ratematrix}
\end{equation} where $k_D$ is the donor fluorescence rate, $k_A$ is the acceptor fluorescence rate,
$k_{ET}$ is the FRET energy transfer rate. The state space is defined as $S = \{D,A,F_D,F_A\}$, where $D$ is the donor position, $A$ is the acceptor position, $F_D$  is the donor fluorescence, and $F_A$ is the acceptor fluorescence. 

Note that if $r=0$, the CTMC is reduced to a two-state system transitioning between states $A$ and $F_A$ with rate $k_A$.

Assuming the dynamic averaging regime, one may easily derive the common time-homogeneous FRET efficiency. Observation of an acceptor photon only occurs when energy transfer occurs, i.e., if we have a transition from $D \rightarrow A$. Let $\tau_D$ and $\tau_{ET}$ be the transfer times of $D\rightarrow F_D$ and $D \rightarrow A$ respectively. Then using Equation \ref{eq:time-inhomogeneousFRET} one obtains,
\begin{align*}
    \mathcal{E}(t) &= \mathbb{P}(\min(\tau_D, \tau_{ET}) = \tau_{ET}) \\& = \frac{\int_0^t k_{ET}(t)}{\int_0^t k_{ET}(t) + k_D t} \\& = \frac{k_D(\frac{R_0}{r})t}{k_D(\frac{R_0}{r})t + k_Dt} \\&= \frac{1}{ \big(\frac{r}{R_0}\big)^6+1}.
\end{align*}

Therefore, the theoretical time-homogeneous FRET efficiency is given by \begin{equation}
    \mathcal{E} = \frac{1}{ \big(\frac{r}{R_0}\big)^6+1}.
    \label{eq:freteff}
\end{equation}
One may approximate this value using two methods: intensity-based FRET and lifetime-based FRET \cite{george,gopich2012theory}. For \emph{intensity-based FRET}, the measurements are counts of observed photons from each dye. Effectively, the experiment measures the probability of success of a binomial random variable with a probability of success $p$ given by the FRET efficiency, $\mathcal{E}$. The best estimator in the absence of experimental corrections is given by the number of successes observed divided by the total number of trials, denoted in Equation~(\ref{eq:EI});\cite{casella2021statistical, Nagy1998, GORDON1998,gopich2012theory, agam2023reliability}
\begin{equation}
    \mathcal{E}_I = \frac{I_A}{I_A + I_D}.
    \label{eq:EI}
\end{equation}

For \emph{lifetime-based FRET}, consider $$\mathcal{E} + P(\min(\tau_D, \tau_{ET}) = \tau_{D})=1.$$ 
Noting that since $P(\tau_D > t| \min(\tau_D, \tau_{ET}) = \tau_{D}) \sim \text{exp}(k_D + k_{ET})$ \cite{Ross}, the FRET efficiency can be calculated in terms of the lifetimes, 
\begin{equation}
    \mathcal{E} = 1 - \frac{\tau_D'}{\tau_D},
    \label{eq:EL}
\end{equation}
where $\tau_D' = (k_D + k_{ET})^{-1}$ is the lifetime of the donor in the presence of the acceptor, and $\tau_D = k_D^{-1}$ is the lifetime in the absence of the acceptor. Hence, the measurements are observed lifetimes and an estimate for the mean lifetime of the donor, $\tau_D$. The FRET efficiency is estimated by approximating the mean, and hence the rate, of this exponential random variable \cite{george}.

\subsection{\label{sec:dynshift}The Dynamic Shift}
    
Consider a sample drawn from a population with a distribution of fluorescence rates $K(x)$ such that the probability of an individual having a specific rate is given by the distribution $\pi(x)$. Then the average lifetime is \begin{equation}
    \overline{\tau} = \mathbb{E}[\tau] = \int_{\mathbb{R}^d} \mathbb{E}[\tau|K(x)]d\pi(x) = \int_{\mathbb{R}^d}\frac{1}{K(x)}d\pi(x).
\end{equation} However, the lifetime resulting from the average rate is given by \begin{equation}
    \underline{\tau} = \frac{1}{\mathbb{E}[K(x)]}=\frac{1}{\int_{\mathbb{R}^d} K(x)d\pi(x)}. 
\end{equation} Therefore, by Jensen's inequality \cite{Billingsley_measure}, using the fact that $\phi(x) = \frac{1}{x}$ is convex for $x\in[0,\infty)$, it must be that \begin{equation}\underline{\tau} = \frac{1}{\mathbb{E}[K(x)]} \leq \mathbb{E}\bigg[\frac{1}{K(x)}\bigg] = \overline{\tau}.\end{equation}  Consequently, the average lifetime for a mixture of states will be greater than that of the associated average state. This phenomenon is known as the dynamic shift.

\begin{figure}[H]
    \centering
    \includegraphics[width = 0.45 \textwidth, height = 0.65\textwidth]{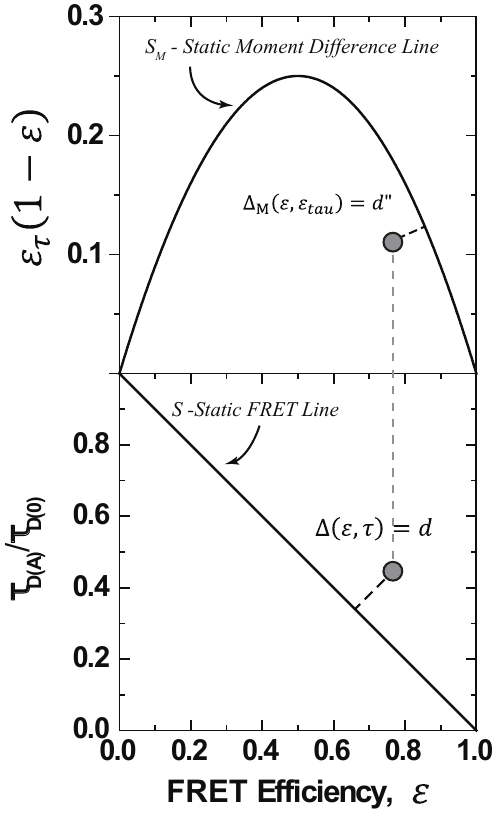}
    \caption{Visualization of the definition of the dynamic shift using normalized values: $S = \{(\mathcal{E},\tau)\in (0,1)^2: \mathcal{E} + \tau - 1 = 0\}$ in the bottom. The top figure is a visualization of the moment difference dynamic shift. For the moment difference we used $\mathcal{E}_{\tau}=1-\frac{\overline{\tau_{D(A)}}}{\tau_{D(0)}}$.}
    \label{fig:Dyn_shft_cartoon}
\end{figure}

We introduce a new quantitative definition of the dynamic shift $\Delta$ for a point $(\mathcal{E}',\tau')$ in the plane, given by the signed distance from the point to the static line, $S = \{(\mathcal{E},\tau): \mathcal{E} = 1 - \tau\}$ as shown in Figure~\ref{fig:Dyn_shft_cartoon}.

One can find an expression for the dynamic shift using standard analytic geometry. The formula in the normalized lifetime case, $\frac{\tau_{D(A)}}{\tau_{D(0)}} = \tau$, is given by \begin{equation}\Delta(\mathcal{E},\tau) = \frac{\mathcal{E}+\tau - 1}{\sqrt{\mathcal{E}^2 + \tau^2}}.\label{eq:dynshift}\end{equation} The definition of the dynamic shift becomes the signed length of the orthogonal projection of the point onto the static line - how much it deviates from the static line. Under the constraint that the FRET - lifetime pair resides within the unit square, this implies that the dynamic shift has extreme values at $\pm \frac{1}{\sqrt{2}}$ at $(1,1)$ and $(0,0)$. This definition provides a means by which each data point from a smFRET experiment may be assigned a dynamic shift value, and the resulting distribution may be examined. The average dynamic shift can be seen as an average deviation from the static line. With two state transitions, this definition agrees with the definition present in \cite{Multistate_1}. Furthermore, when the average dynamic shift is $0$, one may use the dynamic shift distribution to quantify shot noise inherent in the measurements. 

Another way to view the dynamic shift introduced in \cite{Multistate_1} is the moment difference approach. In this method, one investigates the behavior of the difference between the first and second moments of the FRET distribution, $\mathbb{E}[\mathcal{E}(1-\mathcal{E}]) = \mathbb{E}[\tau(1-\tau)]$. In this way, the effects of multiple states are linearized, while the static line is nonlinear. In this case, the dynamic shift can be seen as a consequence of Jensen's inequality but for concave functions. When dynamic mixing is present in the sample, the moment difference should fall below the static line of $\mathcal{E}_{\tau}(1-\mathcal{E})$. Note that when this difference is negative, it implies that the covariance between $\mathcal{E}_{\tau}$ and $\mathcal{E}$ is larger than the average of $\mathcal{E}$. This can occur from shot noise or when the lifetime distribution has a large variance but maintains the same mean. Conditions for this to occur are discussed in Section~\ref{sec:kappa_dyn} To define the dynamic shift from the static moment difference line, one again takes the distance from the point to the static line. The vector between the point and the static line with a length equal to the moment difference dynamic shift will be orthogonal to the tangent line of the static line at the point closest to the point.

The dynamic shift introduced in \cite{Multistate_1} considers an underlying distribution dependent on two separate states. Consider two FRET efficiency states denoted by $\mathcal{E}_i,~i=1,2$ with equal transition rate between the states $\lambda$ for simplicity. Such a two-state system provides valuable insight into the nature of the dynamic shift. When two states are separated on long time scales, $\lambda << 1$, the dynamic shift is slight due to the small amount of mixing during a burst or sample. As the two states mix, corresponding to an increase in $\lambda$, an arc forms between the static FRET-lifetime coordinates, following $(1-\mathcal{E}_1 - \mathcal{E}_2)\mathcal{E} - \mathcal{E}_1\mathcal{E}_2$. As $\lambda \rightarrow \infty$, this process culminates in a point mass FRET-lifetime distribution with a dynamic shift at the maximum of this arc. Therefore, the dynamic shift can be seen as a metric of the amount of mixing between states. Two-state transition systems can be used to understand the transition rates between stable states in biomolecules conformational dynamics. For the current purpose, it provides a convenient method for interpreting the dynamic shift induced by the dyes. The dynamic shift will most readily be present when there are mixing states, and understanding the influence of mixing between an uncountable number of states is of current interest. It will be shown in Section \ref{sec:results} that the dynamic shift induced by dye dynamics can be viewed as a consequence of the fluctuations in the energy transfer rate during the FRET process. Under common circumstances, the energy transfer rate can be approximated by a two-state system corresponding to the modes of the distribution, essentially leading to a quickly transitioning two-state system.    

\section{\label{sec:methods}Stochastic Models of Fluorescence Dynamics} 

\begin{figure}[H]
    \centering
    \includegraphics[width=.5\textwidth]{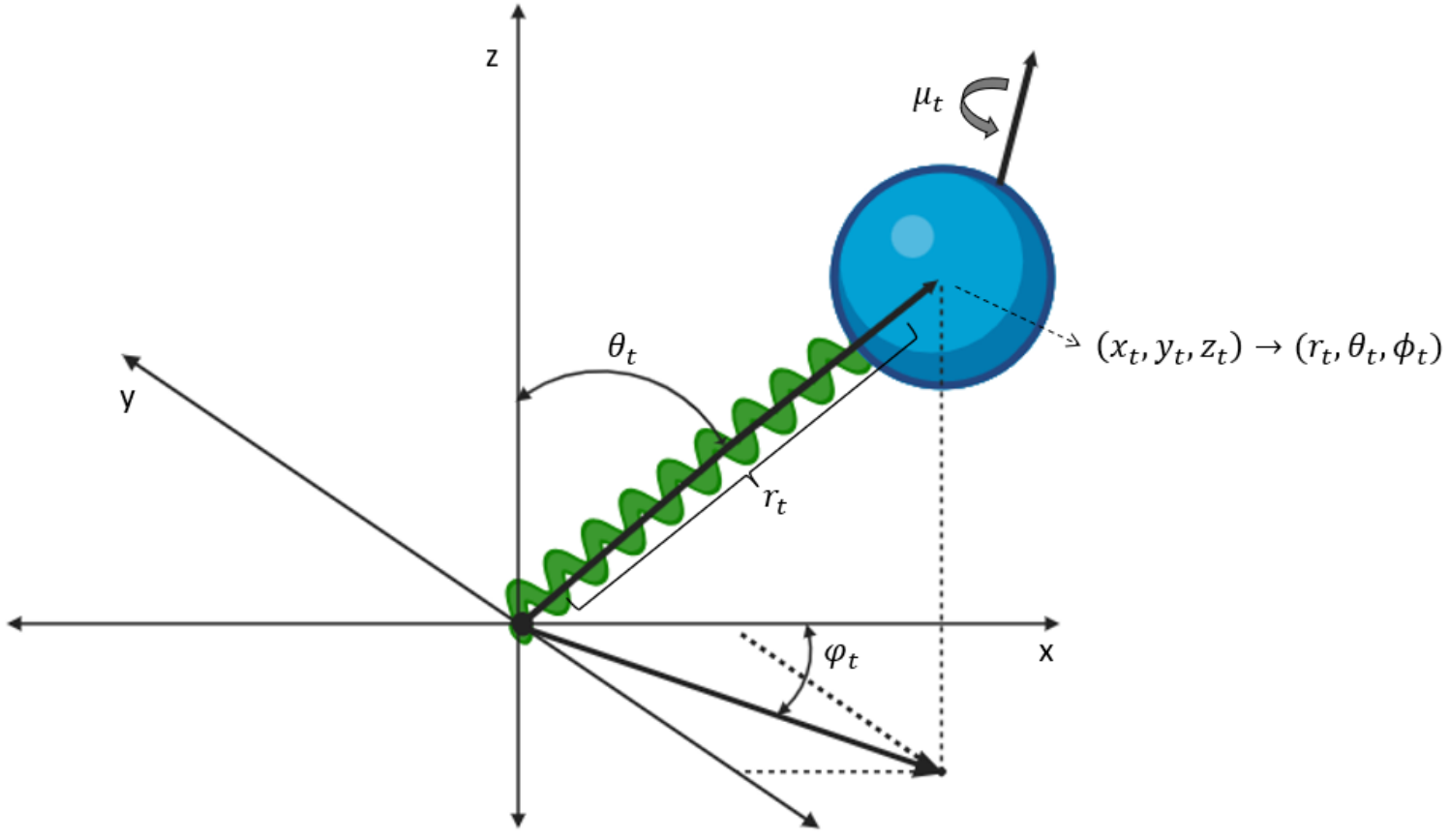}
    \caption{Cartoon showing the coordinate references for the processes. The translational process is expressed in both Cartesian and polar coordinates with $\phi_t$ the azimuthal coordinate, $\theta_t$ the polar coordinate, $r_t$ the radial coordinate, $(x_t,y_t,z_t)$ standard Cartesian coordinates, $\mu_t$ the dipole orientational process 
}
    \label{fig:Dye_config}
\end{figure}

This section presents several models of stochastic fluorescence dynamics related to the smFRET dynamic shift and associated molecular probes. In this way, estimation of the mixture of states, $\pi(x)$ as seen in Section~\ref{sec:dynshift}, is accomplished. Figure \ref{fig:Dye_config} shows the basic coordinate expression for the dye motion. 

Throughout, the dynamics are assumed to evolve on different timescales. Letting $T_P$, $T_D$, $T_O$ represent the timescales of biomolecules dynamics, dye translational dynamics, and dipole orientational dynamics. The order of timescale separation assumed in this work is given by $T_P >> T_D >> T_O$. Further, as in \cite{eilert2018complete,WalczewskaSzewcEtAL,DimuraEtAl,PeulenEtAl,LernerEtAl}, it is assumed that the orientation process and the translational process are independent processes. Note that this is an extremely common assumption since the independence of $\kappa$ and $r$ dynamics is implicitly assumed whenever the average $\kappa$ value is used and whenever static $\kappa$ distributions are employed. \cite{vanbeek2007fretting} Moreover, to provide a clear and succinct picture of the influence of dye dynamics on FRET measurements, the timescale $T_P$ is not considered in the current discussion. However, an extension of this analysis to include this timescale is in development.

\subsection{\label{sec:spring}Spring Models}
The simplest possible model to describe a stationary mean-reverting process is an Ornstein - Uhlenbeck (OU) process.\cite{karatzas1988brownian} This physically represents an overdamped harmonic oscillator subject to noise.\cite{pottier2009nonequilibrium} The OU process is a Gauss-Markov process and, therefore, provides a simple model for thermal fluctuations of the fluorescent dyes. The equation of motion for the state vector $\mathbf{X}_t \in \mathbb{R}^3$ is given by the stochastic differential equation,
\begin{equation}
    d\mathbf{X}_t = K(\mathbf{X}_t -\mathbf{X}_{eq}) dt + \sigma \mathbb{I}\circ dB_t,
\end{equation}
where $K = [k_{i,j}]_{i,j=1}^3$ is a matrix of spring constants and $\mathbb{I}$ is the identity matrix. The notation $\circ dB_t$ denotes the use of Stratonovich integration \cite{Oskendal}, where $B_t$ is Brownian motion. $\sigma >0$ is the volatility of the random fluctuations that are modeled as Brownian motions. We refer to systems such that the spring matrices can be written in the form $K = k\mathbb{I}_{3\times 3}$, as isotropic springs. Otherwise, the system is called anisotropic. 

Both isotropic and anisotropic spring systems with a diagonal spring matrix are considered. The spring coefficients are calculated using the linker chemistry. Utilizing the vibrational frequency of a $C-C$ bond, we find that the spring constant for a single $C-C$ bond is $k = 1010 N/nm$.\cite{hsu1997infrared} Therefore, a system of $N\geq 1$, $C-C$ links is treated as a system of springs in series. Therefore $$\frac{1}{k_{eff}} = \sum_{i=1}^N \frac{1}{k} \rightarrow k_{eff} = k\frac{k}{N}.$$ Finally, to find the length of the linker, we investigate the equilibrium bond length, $L$, in a $C-C-C$ link. Using the law of cosines, we find that $2L = \sqrt{2l^2-2l^2\cos{\theta}}$ with $l$ being the length of a $C-C$ bond. Therefore, the effective length in the linker for each link can be calculated using $l = 1.54 \AA$ and $\theta = 109.5^\circ$.

\begin{figure*}
    \centering
    \includegraphics[scale=0.85]{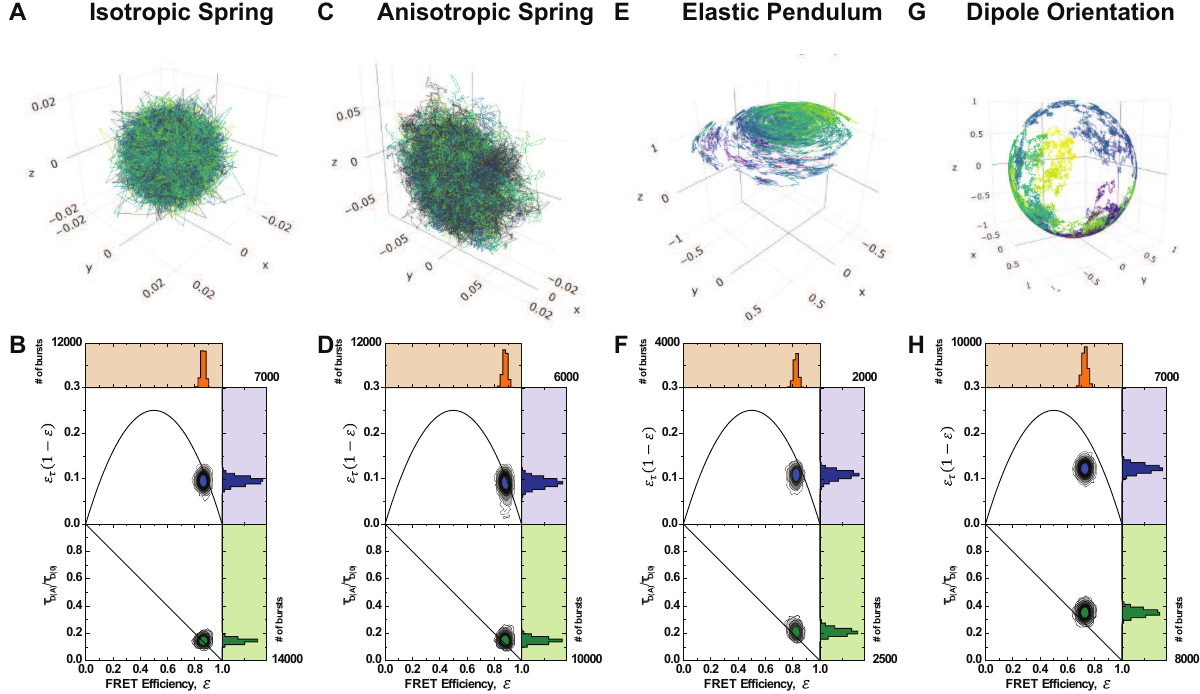}
    \caption{Dye model trajectories and resulting FRET efficiency vs normalized mean fluorescence lifetime and moments difference for A-B) Isotropic Spring, C-D) Anisotropic Spring, E-F) Elastic Pendulum, as described in Section \ref{sec:methods}. G-H) Sample trajectory of a spherical Brownian motion on the unit sphere $S^2$. Such processes are used to model the diffusion of the electric dipole moment. Each of the plots B,D,F,H shows the FRET efficiency vs normalized mean fluorescence lifetime, and the moments difference when the dipole orientation for donor and acceptor is included in the FRET process. Each color in A,C,E,G is an individual burst trajectory. Each case simulates 25000 trajectories over 7 hours. The values in plots B,D,F,H correspond to each simulated state.}
    \label{fig:dyemodels}
\end{figure*}

In the isotropic case, illustrated in Figure \ref{fig:dyemodels}A, the spring matrix is given by $k_{eff}\mathbb{I}_{3\times 3}$. This provides a symmetric three-dimensional Gaussian as the stationary distribution for the isotropic spring.\cite{karatzas1988brownian} It can be seen in \cite{pottier2009nonequilibrium,lavenda2019nonequilibrium} that the variance of this distribution will be given by $\Sigma = \frac{\sigma}{k_{eff}}\mathbb{I}_{3\times 3}.$ 

In the anisotropic case, illustrated in Figure \ref{fig:dyemodels}C, we use a diagonal spring matrix with two entries being $pk_{eff}$ and the third being $k_{eff}$ with $p \in [0,1]$. Therefore, the stationary distribution is an ellipse with major axes determined by the entries of the spring matrix. In this section rotational dynamics have not yet been considered; it will be covered in Section \ref{sec:orientation}.

The two-dimensional dynamics in the anisotropic case can be used to investigate the influence of the orientation of the stationary distribution on the resulting dynamic shift. Such a scenario is exemplified in the case when the planes formed by the major axes of each stationary ellipse are mutually orthogonal. Since the stationary distribution for the isotropic case is a sphere and is perfectly symmetric, this can only arise in the anisotropic case. 

Furthermore, these models have the added benefit of having an analytical expression for the inter-dye displacement, especially in the isotropic case. Since the coordinates will be Gaussian distributed the distance between them is simply Rayleigh distributed. \cite{casella2021statistical} This distribution is unimodal, and therefore the only mixing present is due to the variance of the stationary distributions. This mixing is  therefore strongly dependent on the flexibility of the dyes. 


\subsection{\label{sec:elastic}Elastic Pendulum Model}

The next model for the dye linker dynamics takes a stochastic geometric mechanics approach. Consider the motion of a rigid body attached to a spring that is free to move in space. This system forms an elastic pendulum.\cite{arnol2013mathematical} The following system of Langevin equations describes the motion of a point mass elastic pendulum system subject to white noise; 

\begin{align}
\begin{cases}
        dr_t &= -k_r(r_t - r_{eq}) + \frac{1}{r_t} dt + \sigma_r\circ dB^r_t \\ d\theta_t &= -k_{\theta} \sin{(\theta_t)} + \frac{\sigma_{\theta}^2}{r_t^2 \tan{\theta_t}} \circ dB^{\theta}_t \\ d\phi_t &= \frac{\sigma_\phi}{r_t\sin{(\theta_t)}} \circ dB^{\phi}_t.
\end{cases}
\end{align}

Similar to Equation (12), $B_t$ is Brownian motion, $\sigma > 0$ is the volatility of the random fluctuations that are modeled as Brownian motions, and $k$ is the spring constant. Note that each superscript/subscript is indexed by each of the three components ($r,\theta,\phi$) explained in the next sentences. Importantly, the system is considered in spherical coordinates. The radial dynamics, $r_t$ evolve according to the spring dynamics explained in Section~\ref{sec:spring}, with slight alterations due to the change of coordinates. The angular parts of the motion are given by the standard nonlinear pendulum force in the polar direction $\theta_t$ and free diffusion in the azimuthal direction $\phi_t$. Figure~\ref{fig:dyemodels}E shows a sample dye trajectory.

The flexibility in the angular components is reminiscent of the wobble in a cone model used in previous investigations\cite{kinosita1977theory}, and angular flexibility can be explained via the angular flexibility of $C-C$ bonds themselves. However, unlike the classical wobble in a cone model, thermal noise and dye linker chemistry drive the dynamics and present a purely stochastic system. Moreover, by varying the parameters used, the system shows various behaviors. 

\begin{figure}[H]
    \centering
    \includegraphics[width=.45\textwidth]{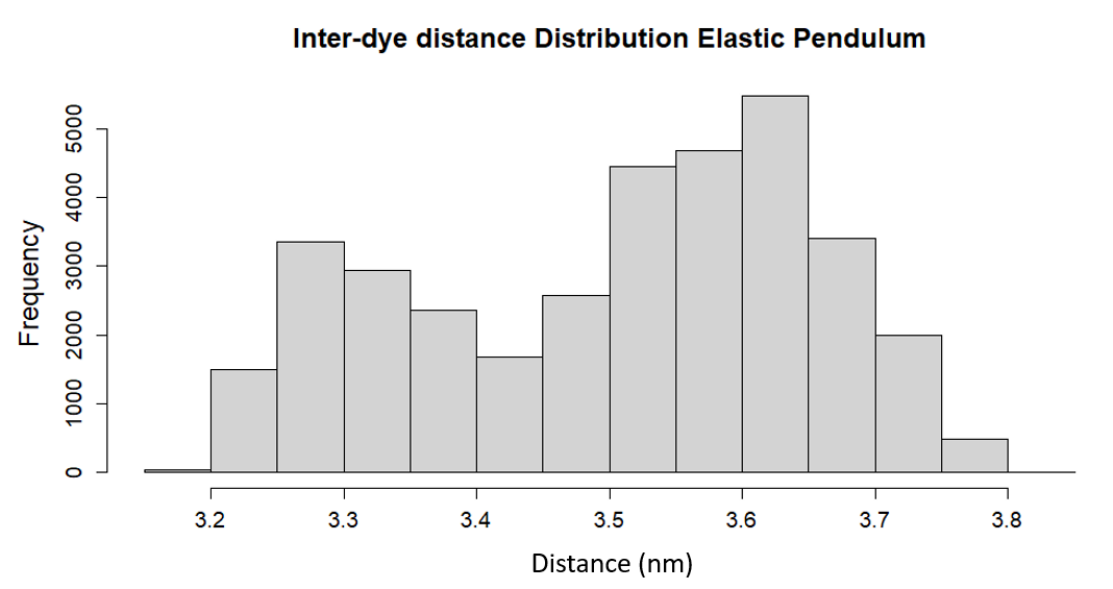}
    \caption{Histogram of inter-dye distances for the elastic pendulum model during an excitation event.}
    \label{fig:dye_bursttime}
\end{figure}

Moreover, this model presents a possible explanation for the dynamic shift induced by dye motion due to the non-Gaussian inter-dye distributions, as shown in Figure~\ref{fig:dye_bursttime}.

This bimodality presents a mixing of two distinct states that are frequently needed to present a dynamic shift. Further, this provides a much larger change in inter-dye displacement than the spring models, which, as mentioned in Section \ref{sec:spring}, do not possess any strong range of translational motion.

\subsection{\label{sec:orientation}Orientational Dynamics}

The final consideration involves the orientational dynamics of the electric dipole moments of the dyes. As discussed in Section \ref{sec:intro}, the F\"{o}rster radius is dependent on the $\kappa$ parameter, which is dependent on the mutual orientations of the electric dipole moments $\hat{\mu}_A, \hat{\mu}_D$ and the inter-dye displacement unit vector $\mathbf{R}$. Typically, $\kappa^2$ is taken as the mean value of $2/3$ when the unit vectors are considered uniformly distributed on the sphere $S^2$.\cite{van2002kappa, vanbeek2007fretting, vandermeer2020kappaphobia} This assumption ignores the temporal aspect of the fluorescent process. Since dyes reorient on timescales faster than fluorescent lifetimes, the energy exchange rate changes during the FRET process. This changes the original CTMC model to a time-inhomogeneous CTMC, and thus, the transfer rates are dependent on the time integral of the infinitesimal transfer rates. \cite{Durrett,Ross, van1992uniformization, stroock2004introduction}  

To incorporate the influence of orientational dynamics on the lifetime distribution and FRET efficiency, consider the dipoles to be fixed to a reference frame of some rigid body with tensor of inertia $\mathbf{I}$. The rigid body of the dye will be subjected to random torques and, therefore will reorient according to the Euler equations\cite{arnol2013mathematical,chirikjian2011stochastic} \begin{align}
    \begin{cases}
        \mathbf{I}d\omega_t + \omega_t \times \mathbf{I}\omega_t = -\nu \omega_t + d\mathbf{W_t} \\ \omega_t = d\mathbf{\Phi_t}
    \end{cases}
\end{align}
where $\omega$ is angular velocity, $\nu$ is the dynamic viscosity of the surrounding fluid, $dW_t$ is a spherical Brownian motion and $\mathbf{\Phi_t}$ is the angular position vector. Assuming the dye is overdamped and hence $d\omega_t = 0$, one obtains the simplified equations \begin{align}
    \begin{cases}
        \omega_t \times \mathbf{I} \omega_t = - \nu \omega_t + d\mathbf{W_t} \\ \omega_t = d\mathbf{\Phi_t}.
    \end{cases}
\end{align}
Making the assumption that the dye is spherical and therefore the inertia tensor may be replaced with a scalar value \cite{arnol2013mathematical} and using the fact that $v\times v = 0$ for any vector $v$ we obtain the simple formula \begin{equation}
    \nu d\mathbf{\Phi_t} = d\mathbf{W_t}
\end{equation} and hence, the dipole diffuses according to a spherical Brownian motion. Spherical Brownian motions can be expressed in terms of the Langevin equations below \cite{Hsu, brillinger2012particle}
 
\begin{align}
\begin{cases}
        d\theta_t &= \frac{\sigma_{\theta}^2}{\tan{(\theta_t)}}dt + \sigma_\theta \circ dB_t \\ d\mathbf{\Phi_t} &= \frac{\sigma_\phi}{\sin{\theta_t}}\circ dB_t
\end{cases}
\end{align}

The rotational diffusion coefficients depend on the hydrodynamic radius of the dye $R_h$ by the classical relation $D = kT/(8\pi \nu R_h^3)$, where $kT$ denotes the product of the Boltzmann constant and the temperature. A sample trajectory is shown in Figure~\ref{fig:dyemodels}G.\\

Note that the stationary distribution for such a system is the uniform distribution, providing an ideal starting stochastic process to test the time-dependent behavior of orientational dynamics.\cite{Oskendal, chirikjian2011stochastic} The key idea is that the excitation of the fluorophores provides a single initial $\kappa$ value. 
The relaxation effects are the object of interest, especially with regard to lifetime duration. The notion that $\kappa$ may be close to $0$ during the entire FRET process for one excitation but higher for another in the same sampling time provides an additional source of variance in the lifetime distribution. Every FRET process can lead to a different equilibrium, which should depend on the rotational diffusion of the dipole moment, with faster reorientation causing an averaging out effect as mentioned in \cite{van2002kappa}. 

\section{\label{sec:results} Sources of Observed Dynamic Shift} 

\subsection{\label{sec:dyeconfig} Dye Configuration}

This section compares the dye models mentioned in Section \ref{sec:methods}. First, inspecting the joint FRET-lifetime distributions using the contour plots and marginal histograms seen in Figure \ref{fig:dyemodels}, one can see the influence of the differing models. These FRET-lifetime distributions were generated by simulating the above-mentioned models in a confocal smFRET environment. As can be seen, the spring models show similar characteristics to the anisotropic model, which shows slightly more dynamic shifting. In addition, the elastic pendulum model produces a noticeable shift where the bulk of the distribution lies off of the static line. Despite the dynamic mixing involved in the purely translational elastic pendulum model, the dynamic shift produced is still not representative of experimental observations. Only through the addition of the orientational motion and time-inhomogenous energy transfer rates does the distribution show the hallmark dynamic shift in both the moment difference as well as direct FRET-lifetime distribution. 

\begin{figure}[H]
    \centering
    \includegraphics[width=.45\textwidth]{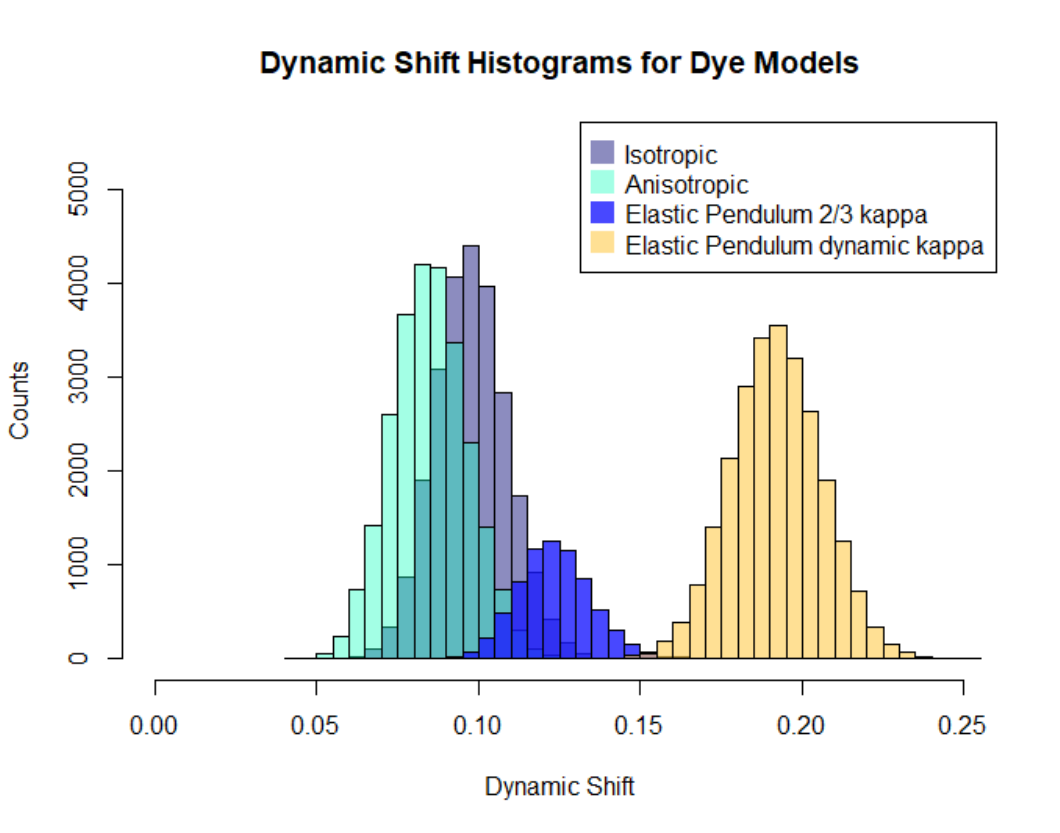}
    \caption{Linear Dynamic shift, $d$, histogram comparison of dye models}
    \label{fig:Dye_Model_Hist}
\end{figure}

The dynamic shift distributions of each dye configuration are calculated using the definition of the dynamic shift shown in Equation (\ref{eq:dynshift}). It has been known from experimental data that the average dynamic shift of dye motion is $\mu(\Delta) \approx 0.2$.\cite{Multistate_1} Using this quantity, the average dynamic shift for the associated models is examined to determine the model that captures the appropriate mean dynamic shift. The comparison of the dynamic shift distributions is shown in Figure \ref{fig:Dye_Model_Hist}. 

In addition, these simulations have no burst noise from background radiation, as this could potentially cloud the impact of the dye motion.\cite{nir2006shot} The noise is solely from the experimental photon loss considerations and the dye motion as dictated by the models and simulation methods. Therefore, the only source of dynamic shifting must be from the dye models.  

\begin{figure*}
    \centering
    \includegraphics[scale=0.85]{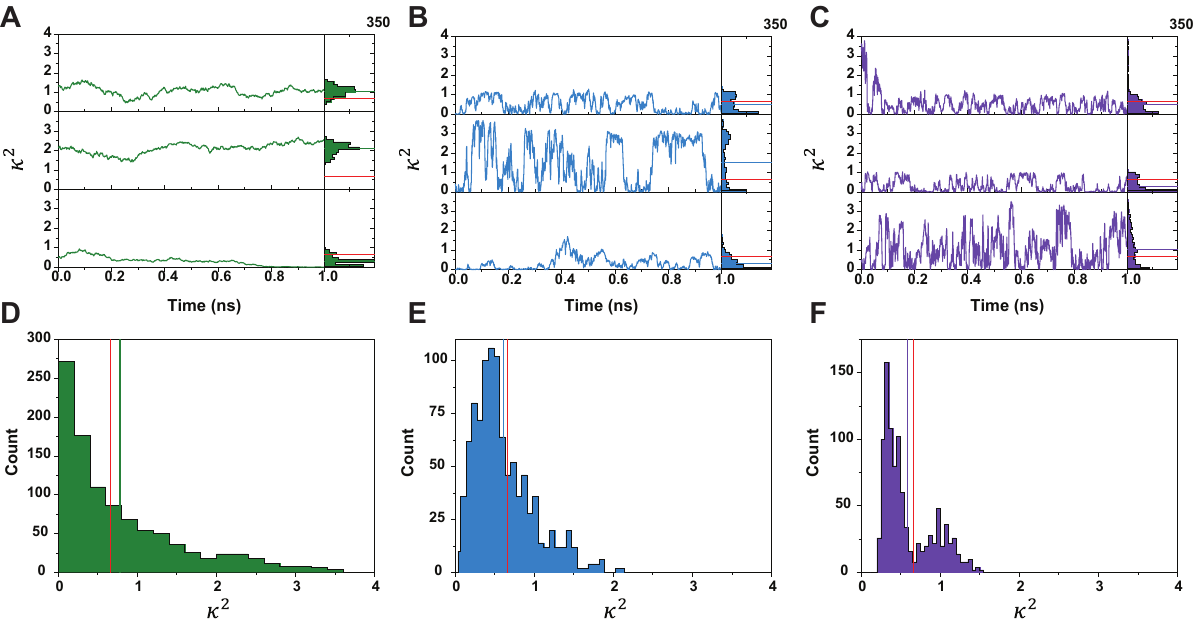}
\caption{Comparison of $\kappa^2$ trajectories for three common cases. Figure $A - C$ shows the sample paths of $\kappa^2$ during an excitation event, these can be seen as four realizations of $\kappa^2$ during the same burst. Each path is representative of a single energy transfer event. Figures $D-F$ show the associated distribution of average $\kappa^2$, the red vertical lines in figures $D-F$ are the isotropic $2/3$ average, where the colored lines are the mean of the path average. 
Figures $A$ and $D$ show a pair of dyes for which the rotational diffusion is only an order of magnitude greater than the translational diffusion of the dye. Figures $B$ and $E$ show the sample behavior when one dye has a rotational diffusion three orders of magnitude greater than translational and one dye one order of magnitude greater. Finally, Figures $C$ and $F$ show the case in which both dyes have rotational diffusions three orders of magnitude greater than translational.}
    \label{fig:Kap_trj}
\end{figure*}

Curiously the dynamic shift densities shown in Figure \ref{fig:Dye_Model_Hist} exhibit similar variances but differing mean dynamic shift values, as seen in Table \ref{tab:dynamic_shift_values}. 
\begin{table}[hbt!]
\caption{\label{tab:dynamic_shift_values}
\textbf{Dynamic Shift:} Mean and standard deviation of the dynamic shift for each model in Figure \ref{fig:Dye_Model_Hist}.}
\begin{tabular}{|c|c|c|}
\hline
     Model & $\mu(\Delta)$ & $\sigma(\Delta)$ \\ \hline
   Isotropic Spring  & 0.10 & 0.01 \\ \hline  Anisotropic Spring & 0.01 & 0.01\\ \hline Elastic Pendulum $\kappa \approx 2/3$ & 0.12 & 0.01  \\ \hline  Elastic Pendulum Dynamic $\kappa$ & 0.20 & 0.10 \\  
\hline
\end{tabular}
\end{table}
While the elastic pendulum model does not completely capture the mean dynamic shift, with the addition of a dynamic $\kappa^2$ value, the proper dynamic shift appears. This is in stark contrast to previous suggestions that the dynamic shift is a result of the accessible volume of the dye. Despite the elastic pendulum having an accessible volume comparable to those found in all-atom molecular dynamic simulations and exhibiting dynamic mixing between two states, the resulting dynamic shift is less than expected. This suggests that the inclusion of time-inhomogeneities in the F\"{o}rster radius due to orientational factors is essential to the description of the dynamic shift.

It is important to note that the spring models can be made to incorporate a dynamic $\kappa^2$ parameter. However, the model parameters based on linker composition prevent these models from being used in earnest, as the accessible volumes are smaller than physically reasonable. This can be seen by investigating the variance of their associated stationary distributions. As mentioned in Section \ref{sec:spring}, the variance along each axis in the isotropic case will be $\sigma/k_{eff}$, or $K_BT/\gamma k_{eff}$, where $\gamma$ is the local friction. This produces a stationary distribution with a $3$ standard deviation radius of less than an Angstrom. 

\subsection{\label{sec:kappa_dyn} $\kappa^2$ Dynamics}

An important consideration brought to attention in Section \ref{sec:dyeconfig} is the influence of $\kappa^2$ dynamics on FRET lifetime pairs. The key issue in incorporating $\kappa^2$ dynamics into FRET uncertainty quantification has been to assess $\kappa^2$ as a stationary object during energy transfer. This is done by considering $\kappa^2$ as being chosen from its equilibrium distribution, assuming that it follows a discrete state Markov chain, or simply using the $2/3$ approximation. \cite{pirchi2016photon,harris2022multi,ingargiola2016fretbursts, eilert2018complete} From there, one can use the mean and standard deviation in uncertainty quantification.\cite{vandermeer2020kappaphobia,van2002kappa, vanbeek2007fretting} However, these approximations ignore the time inhomogeneous nature of the FRET process, as mentioned in Section \ref{sec:fret}. Notably, the probability of a donor fluorescence event, as shown in Section \ref{sec:fret}, is dependent on the integral of the energy transfer rates as well as the lifetime distribution. \cite{Durrett,dobrow2016introduction, stroock2004introduction, van1992uniformization} While the use of an ergodic approximation might mitigate these concerns, this relies on the convergence of the long run time average to the long run spatial average. The short time span involved in the FRET process, especially in slow rotational diffusion regimes, should not be considered long enough to invoke such an approximation. Furthermore, this mitigation ignores the dependence of the energy transfer rate and FRET efficiency on the history of the $\kappa^2$ process. It is important to note that experimental bursts are one-millisecond in length, equivalent to one molecule. The equivalent measurement used in the simulations is one burst per second, leading to approximately 25,000 bursts. As seen in Section \ref{sec:fret}, the energy transfer rate is dependent on the integral of the $\kappa^2$ path. In Figure \ref{fig:Kap_trj} we keep track of the $\kappa^2$ value and display it's stochasticity. For example, in Figures \ref{fig:Kap_trj}A and B, the trajectories have similar mean values after the $1ns$ run.
However, if one were to investigate the energy transfer rate at time $0.5$ one would find that the probability of energy transfer is much lower for path B rather than being similar to path A. One can see how this integral fluctuation may be reduced in the presence of fast dyes. As shown in Figure \ref{fig:Kap_trj}C, the rapid oscillations induced by the swift rotational diffusion of the dyes produce highly clustered $\kappa^2$ trajectories. However, as can be seen in the associated average $\kappa^2$ distribution in Figure \ref{fig:Kap_trj}F, the resulting average distribution becomes bimodal. This bimodality can be seen in the trajectories in which the oscillations can vary between smaller values or larger values. Notably, this is heavily dependent on the radial-dipole dot product term in the definition of $\kappa^2$. Furthermore, with the decrease in rotational diffusion, the $\kappa^2$ process predictably oscillates much slower. This yields trajectories such that larger values are maintained for longer. These larger stretches of high $\kappa^2$ values can lead to much shorter donor lifetimes, but similarly, this also leads to longer stretches of small $\kappa^2$ values. Overall, slower rotational diffusion will show a larger temporal correlation. Hence, the stationary distribution will play a much more important role, since the $\kappa^2$ trajectory is not as likely to change as drastically from its initial value.

The skew of the mean $\kappa^2$ value shows the flaw in the averaging assumption. While the mean $\kappa^2$ average is indeed $2/3$, the most common values for the mean $\kappa^2$ are below this value. In fact, one can see that out of the three distributions of average $\kappa^2$ the only one in which the most common value is $2/3$ is Figure \ref{fig:Kap_trj}E, wherein as mentioned above, the temporal fluctuations of the $\kappa^2$ path will play a larger role due to their larger time heterogeneity than the paths in Figure \ref{fig:Kap_trj}A and \ref{fig:Kap_trj}C. 

The dynamic shift produced by dye motion is then a product of the variability of the $\kappa^2$ trajectories. This temporal heterogeneity \emph{during} the energy exchange process provides the crucial mixing element to significantly slow the the average sampled donor lifetime. While it can be seen in Figure \ref{fig:dyemodels}
that inter-dye radial displacement dynamics can produce a slight dynamic shift, the mixing is not strong enough to produce the lifting seen by considering the direct change in energy transfer rate caused by rotational dynamics. Figure \ref{fig:Kap_trj}D- \ref{fig:Kap_trj}E can be seen as representative of the average $\kappa^2$ values during a single burst. If one simplifies the path-based description above to these distributions, it is simple to see that in the case of two rapidly rotating dyes, the bimodality of the distribution provides two populations of energy transfer rate. The switching between the two during the burst provides a direct comparison to the two-state systems investigated in \cite{Multistate_1,Multistate_2}. However, it must be stressed that the path-by-path heterogeneity of the FRET process is crucial to the analysis. While the average values shown can be representative of normalized time integrals, one must consider that the plots shown in Figure \ref{fig:Kap_trj} each last for one nanosecond, and therefore do not capture the stopping time of fluorescence or energy transfer. As noted, the time-dependent FRET efficiency $\mathcal{E}(t)$ will vary path-by-path, producing the observed lifetime distribution.

\section{\label{sec:discussion} Discussion} 

In this work, we showed how dynamic shift results from dye motion. By presenting the first physics-based model for fluorescence dynamics, incorporating dye linker chemistry and fluorescent dye composition, it has become clear that the time-inhomogeneous nature of the F\"{o}rster radius is an essential part of the dynamic shift. This is shown by noting that models with proper accessible volumes do not demonstrate experimentally observed dynamic shifts and noting that the time-inhomogeneous nature of the FRET process depends on the path space dynamics of $\kappa^2$ trajectories. Furthermore, it was shown that the dynamics of $\kappa^2$ trajectories change with the selection of fluorescent dyes used, with common situations such as an organic dye paired with a fluorescent biomolecules exhibiting noticeable differences from a pair of organic dyes. While the average $\kappa^2$ stays consistent with the $2/3$ isotropic assumption, the fluctuations during the FRET process cause fluctuations in the FRET-lifetime distribution during the burst. These fluctuations provide a source of dynamic mixing and, hence, a dynamic shift. In short, FRET efficiency does not change, but the lifetime does.\\

In this work the results are for only spherical dyes, however, the rotational Langevin equations used in section \ref{sec:orientation} can be utilized for arbitrary intertia tensors. While this is simple to state, the resulting equations become arbitrarily difficult to work with, as well as simulate. Non-spherical dyes require different intertia tensors and rotational processes, which produce unique $\kappa^2$ trajectories.

An additional consideration for future work is the coupling between translational motion and orientational motion. While this is a common assumption, as it is invoked each time dynamic averaging is used for $\kappa^2$ and is used in other theoretical analyses such as in \cite{eilert2018complete}, it is an important direction of consideration. As noted in \cite{vanbeek2007fretting} there is a marked correlation between translational motion and $\kappa^2$. This can be further emphasized by the considerations of Sections \ref{sec:elastic} and \ref{sec:orientation} wherein one may couple the orientational rigid body dynamics and the dynamics of the elastic pendulum. Such a system is known to have non-trivial coupling in the classical case that is likely to remain in the overdamped case. These considerations, however, increase the difficulty of analysis and simulation considerably as one must consider the dynamics of a full state space model on the non-compact, non-abelian Euclidean group $SE(3)$ \cite{chirikjian2011stochastic} rather than independent processes as used here and in \cite{eilert2018complete}.    \\  

Furthermore, the work shown here provides an important framework for analyzing other sources of the dynamic shift. The algorithms used to simulate the experiments can easily be scaled to incorporate another timescale. This opens the door to studying the influence of biomolecules dynamics on the dynamic shift. By using reaction coordinate Langevin models, one may investigate the influence of differing energy landscapes on the dynamic shift, as well as the influence of a more dynamic inter-dye distance vector. This can be done swiftly due to the low computational cost incurred by the stochastic simulation of the FRET process. Additionally, by introducing a new form for the dynamic shift, distributional quantities may be accessed more readily, providing a new tool to analyze the motion of underlying molecules of interest as well as dye dynamics. Additionally, while our simulated experiments were conducted in the confocal environment the results carry over to TIRF measurements as well. 

In conclusion, this work has demonstrated the importance of time-inhomogeneities in the FRET process and their influence on resulting measurements. The path dependence of the FRET efficiency, as well as the influence of a time-varying F\"{o}rster radius during the FRET process, is shown to be non-negligible. In order to conduct optimal uncertainty quantification for smFRET measurements the path dynamics of $\kappa^2$, not just the average values, must first be understood. Therefore, the anisotropic behaviors of fluorescent dyes are of vital importance. By gaining an understanding of the orientational dynamics of fluorescent dyes, better uncertainty quantification for FRET measurements can be done. Additionally, this relationship between the full state space dynamics of a fluorescent dye and resulting smFRET measurements emphasizes the need for more stochastic geometric mechanical considerations in fluorescent measurements and biology. 

\section{\label{sec:acknowledgements} Acknowledgement} 
The authors acknowledge the Palmetto Cluster at Clemson University. This work was partially supported by NIH grants 1R15AI137979, and 1R15CA280699 and NSF CAREER MCB 1749778 to HS.

\section{Appendix\label{sec:appendix}}

\subsection{Table of Parameters for Figure 7}


\begin{table}[hbt!]
\caption{\label{tab:Figure7Table}Simulation parameters for Figure \ref{fig:Kap_trj}.}
\begin{tabular}{|c|c|c|c|}
\hline
    Parameter & Plots A \& D& Plots B \& E & Plots C \& F  \\ \hline  
     Donor Rotational Diffusion & 15 nm$^2$/s & 15 nm$^2$/s & 200 nm$^2$/s\\ \hline 
    Acceptor Rotational Diffusion & 15 nm$^2$/s & 150 nm$^2$/s & 200 nm$^2$/s\\ \hline 
    Translational Rotation & 0 nm$^2$/s & 0 nm$^2$/s & 0 nm$^2$/s\\
    
\hline
\end{tabular}
\end{table}

\bibliography{FRET_Ref}

\end{document}